\documentclass[12pt]{article}
\usepackage{amsmath,epsf,amssymb,latexsym,cite,setspace}
\usepackage[ps,dvips,matrix,arrow,frame,import,curve,color]{xy}
\xyoption{ps}
\xyoption{dvips}

\newif\iffigs\figstrue

\DeclareFontFamily{U}{rsf}{}
\DeclareFontShape{U}{rsf}{m}{n}{
  <5> <6> rsfs5 <7> <8> <9> rsfs7 <10-> rsfs10}{}
\DeclareMathAlphabet\Scr{U}{rsf}{m}{n}

\def\pplogo{\vbox{\kern-\headheight\kern-29pt
\halign{##&##\hfil\cr&{%\sc
\ppnumber}\cr\rule{0pt}{2.5ex}&\ppdate\cr}
}}
\makeatletter
\def\ps@firstpage{\ps@empty \def\@oddhead{\hss\pplogo}%
  \let\@evenhead\@oddhead % in case an article starts on a left-hand page
}
\def\maketitle{\par
 \begingroup
 \def\thefootnote{\fnsymbol{footnote}}
 \def\@makefnmark{\hbox{$^{\@thefnmark}$\hss}}
 \if@twocolumn
 \twocolumn[\@maketitle]
 \else \newpage
 \global\@topnum\z@ \@maketitle \fi\thispagestyle{firstpage}\@thanks
 \endgroup
 \setcounter{footnote}{0}
 \let\maketitle\relax
 \let\@maketitle\relax
 \gdef\@thanks{}\gdef\@author{}\gdef\@title{}\let\thanks\relax}
\makeatother

\def\cH{\Scr{H}}
\def\Z{\Scr{H}}
\def\cHb{{\overline{\Scr{H}}}}
\def\Hb{\overline{H}}
\def\Ht{\widetilde{H}}
\def\C{{\mathbb C}}

\def\P{{\mathbb P}}
\def\R{{\mathbb R}}
\def\Z{{\mathbb Z}}

\def\Tr{\operatorname{Tr}}

\def\Res{\operatorname{Res}}

\def\GU{\operatorname{U{}}}

\def\rank{\operatorname{rank}}

\def\p{\partial}

\def\cL{{\Scr L}}

\def\ff#1#2{{\textstyle\frac{#1}{#2}}}

\def\ep{\epsilon}

\def\nv{\mathbf{n}}

\def\Qv{\mathbf{Q}}
\def\Qc{\mathcal{Q}}
\def\rv{\mathbf{r}}
\def\K{\mathcal{K}}

\def\N{\mathcal{N}}

\def\la{\langle}
\def\ra{\rangle}

\def\lad{\langle\!\langle}
\def\rad{\rangle\!\rangle}

\def\tp{{\theta^+}}

\def\tbm{\overline{\theta}^-}

\def\Phib{\overline{\Phi}}
\def\sigmab{\overline{\sigma}}
\def\psib{\overline{\psi}}
\def\sigmah{\hat{\sigma}}

\def\omegah{\hat{\omega}}
\def\Sigmab{\overline{\Sigma}}
\def\lambdab{\overline{\lambda}}

\def\Wt{{\widetilde{W}}}
\def\Wte{{\widetilde{W}_{\text{eff}}}}

\def\Orbt{\C^3/\Z_{(2N+1)(2,2,1)}}

\def\xib{\overline{\xi}}
\def\chib{\overline{\chi}}
\def\etab{\overline{\eta}}
\def\rhob{\overline{\rho}}
\def\zb{\overline{z}}

\begin{document}
\setcounter{page}0
\def\ppnumber{\vbox{% \baselineskip 25pt
\vspace{15mm}
\hbox{DUKE-CGTP-05-07}
\vspace{2mm}
\hbox{hep-th/0507187}}}
\vspace{2mm}
\def\ppdate{\today} \date{}

\title{\LARGE $A$-Model Correlators from the Coulomb Branch  \\[10mm]}
\author{
Ilarion V. Melnikov and M. Ronen Plesser\\[3mm]
\normalsize Center for Geometry and Theoretical Physics \\
\normalsize Box 90318 \\
\normalsize Duke University \\
\normalsize Durham, NC 27708-0318
}

{\hfuzz=10cm\maketitle}

\def\Large{\large}
\def\LARGE{\large\bf}

\vskip 1cm

\begin{abstract}
We compute the contribution of discrete Coulomb vacua to $A$-Model correlators in toric Gauged Linear
Sigma Models.  For models corresponding to a compact variety, this determines the correlators at arbitrary
genus.  For non-compact examples, our results imply the surprising conclusion that the quantum cohomology
relations break down for a subset of the correlators.
\end{abstract}

\vfil\break

%%%%%%%%%%%%%%%%%%%%%%%%%%%%%%%%%%%%%%%%%%%%%%%%%%%%%%%%%%%%%%%%

\section{Introduction}  
Topological Field Theory (TFT) is a powerful tool for studying the RG-invariant
properties of non-trivial quantum field theories.  A particularly important
class of examples is provided by the $A$ and $B$ twists of an $\N=(2,2)$ SUSY 
Non-Linear Sigma Model (NLSM) defined on a Riemann surface $\Sigma$ \cite{W:topsig,W:mirtop}.
These TFTs provide rich examples of solvable quantum field theories, and they have
important applications to compactification in string theory. In
addition, these TFTs can be used to study enumerative geometry and 
refined topological invariants, such as the Gromov-Witten invariants, 
of many manifolds.  Finally, these theories provide a natural setting for the study
of mirror symmetry of Calabi-Yau manifolds\cite{W:mirtop}.  Although the TFT 
perspective on the NLSM is immediately conceptually useful, aside from particularly 
tractable examples such as the classic work of Candelas {\it et al} \cite{COGP} and 
its various generalizations\cite{COFKM:2param1,COFKM:2param2,HKTY:hyper,HKTY:inter}, direct study 
of these models is difficult.

Remarkably, a large class of NLSMs, including those corresponding to 
Calabi-Yau manifolds constructed as hypersurfaces or complete intersections in
Fano toric varieties, may be constructed as IR limits of certain $\N=(2,2)$ SUSY
abelian gauge theories termed Gauged Linear Sigma Models (GLSMs)\cite{W:phases}.  
The RG-invariant observables of the $A$ and $B$ models may be computed in the
massive theory.  The $A$-model in particular has been studied extensively in 
\cite{W:phases,MP:summing}.  It was found that many properties of the GLSM, 
including the correlators in the topologically twisted models, are constrained by toric 
geometry.  For example, toric methods allow an explicit and general formulation of the 
instanton sum for the ``toric'' subset of  $A$-model correlators\cite{MP:summing}.  The 
study of these instanton sums is enlightening:  it allows for a careful definition of 
the monomial-divisor mirror map, and it reduces mirror symmetry to the mirror map, 
a non-trivial renormalization of the GLSM versus NLSM parameters. 
Furthermore, for a genus zero Riemann surface, the {\em quantum restriction
formula} of \cite{MP:summing} reduces the computation of topological correlators for a 
Calabi-Yau hypersurface to the computation of correlators on the ambient Fano toric variety.
Where the mirror map is known, this allows for an explicit verification of 
mirror symmetry at the level of TFT.  These are powerful results.  Still, the 
instanton sums are unwieldy.  In general they may be quite intricate, and simple but 
important properties of the correlators, like the quantum cohomology relations, seem to 
follow from obscure relations between intersection numbers on various toric 
varieties \cite{MP:summing,ME:glsmtach}.  Finally, the toric methods for performing the
gauge instanton sums have only been developed for genus zero correlators, and their extension
to $g>0$ correlators is non-trivial.

In this work we reduce the computation of $A$-model correlators in a wide class of GLSMs to
a well-studied algebraic problem.  Our result is simple to state.  Consider a GLSM with
a set of chiral matter $\N=(2,2)$ multiplets $\Phi^i$, $i=1,\ldots,n$,  charged under 
a gauge group $\left[\GU(1)\right]^r$ with charges $Q^a_i$, $a=1,\ldots,r$, and zero superpotential for
the matter fields.  Such models will be referred to as {\em toric} GLSMs.  Upon twisting,
the local $A$-model observables are found to be functions of the $\sigma_a$, the bosonic 
super-partners of the gauge fields.  The most general $A$-model correlator on $\Sigma_g$, a Riemann 
surface of genus $g$, may be obtained from linear combinations of 
$\la \sigma_{a_1} (z_1) \cdots \sigma_{a_s} (z_s) \ra_g$ and derivatives of these with respect to the 
GLSM parameters.  Since this is a TFT computation, the correlator is independent of the 
{\em generic} points $z_1,\ldots,z_s$ on $\Sigma_g$.  Following a standard notation, we will 
denote these correlators by $\la F(\sigma)\ra_g$, where $F(\sigma)$ is to be understood as a 
power series in $\sigma_{a_k}(z_k)$ with a generic choice of the $z_{k}$.  In a sense made more precise 
below, ``most'' toric GLSMs possess a region of parameter space where the theory has a 
number of discrete Coulomb vacua determined as solutions to the equations of motion for a 
certain effective twisted superpotential $\Wte(\sigma)$.  This class includes all GLSMs 
corresponding to compact toric varieties.  For these {\em compact, toric} GLSMs there is an 
additional simplification:  there exists a region of the parameter space where these discrete 
Coulomb vacua are the {\em only} vacua.  In this case, the $A$-model correlators at genus $g$ 
are given by
\begin{equation}
\label{eq:corresult}
 \la F(\sigma) \ra_g = 
   \sum_{\sigmah| d\Wte(\sigmah) = 0} \Ht(\sigmah)^{g-1} F(\sigmah),
\end{equation}
where $\Ht(\sigma) = H \prod_{i=1}^n \xi_i $, $H$ is the Hessian of $\Wte$, and 
$\xi_i = \sum_a Q^a_i \sigma_a.$  The equations of motion that follow from $d\Wte=0$ are 
polynomial in the $\sigma_a$, so the computation of $A$-model correlators is now 
reduced to an algebraic problem.

This form is convenient for obtaining explicit correlators in many toric examples. Furthermore,
by the use of the quantum restriction formula of \cite{MP:summing}, it becomes a useful
tool to compute genus zero $A$-model correlators on GLSMs corresponding to Calabi-Yau surfaces.
In addition, eqn.(\ref{eq:corresult}) manifestly satisfies the quantum cohomology relations,
and, as we will see in more detail below, it is a useful probe for the physics of the 
Coulomb branch of the GLSM. 

Since we have not coupled the TFT to topological gravity, our
result is of limited use for the computation of general Gromov-Witten invariants---factorization
of the correlators implies that our higher genus results do not generate ``new'' invariants 
for $g>0$.  However, we believe that for the purposes of enumerative geometry, eqn.(\ref{eq:corresult})
is a neat packaging of the requisite combinatorics. Essentially, while it is true that the $g>0$ 
correlators may be obtained by factorization from the $g=0$ correlators, if one interested in
explicit numbers, eqn.(\ref{eq:corresult}) may eliminate much algebraic suffering.

It is no accident that the form we find is reminiscent of the correlators in topological 
Landau-Ginzburg models studied by Vafa \cite{VAFA:tlg}.  In fact, our analysis is a 
simple extension of those techniques to include the zero modes of the matter fields.  These
additional zero modes are the source of the factor of  $\prod_i \xi_i$ in our expression.  
We will discuss this further below.

While eqn.(\ref{eq:corresult}) computes the correlators in compact toric GLSMs, in non-compact toric 
GLSMs the above expression is only a part of the story.  In general, only a subset of the solutions
to $d\Wte=0$ correspond to Coulomb vacua, and eqn.(\ref{eq:corresult}) provides the correct measure
for the contributions to the correlators due to these vacua.  In addition, other, non-Coulomb vacua 
may also contribute, and these contributions may invalidate the quantum cohomology relations.

The rest of this note is organized as follows.  In section \ref{s:over} we briefly review 
the toric GLSM and the corresponding $A$-model, and we take care to distinguish the GLSM phases
according to the properties of the Coulomb vacua.  In section \ref{s:proof} we prove eqn.(\ref{eq:corresult})
by studying the $A$-model localization onto the Coulomb vacua.  We provide 
some applications of the result to compact toric GLSMs in section \ref{s:examples}.  
In section \ref{s:zforreal} we turn to a study of a non-compact example, and we conclude with 
a discussion in section \ref{s:discuss}.
%%%%%%%%%%%%%%%%%%%%%%%%%%%%%%%%%%%%%%%%%%%%%%%%%%%%%%%%%%%%%%%%
%%%%%%%%%%%%%%%%%%%%%%%%%%%%%%%%%%%%%%%%%%%%%%%%%%%%%%%%%%%%%%%%
\section{A GLSM Overview} \label{s:over}

\subsection{Some Superspace Details}
The GLSM is a $d=2$ abelian gauge theory with $\N=(2,2)$ supersymmetry.  The 
field content is neatly summarized in terms of $\N=(2,2)$ multiplets.  The 
matter fields belong to chiral multiplets 
$\Phi^i = (\phi^i,\psi^i_\pm, \psib^i_\pm, F^i)$,
with $\phi^i$ a complex scalar, $\psi^i_\pm$ left/right-moving Weyl fermions,
$F^i$ a complex auxiliary field, and $i=1,\ldots,n$.  These fields are charged 
under the gauge group $G=\left[\GU(1)\right]^r$ with integral charges $Q^a_i$, 
$a=1,\ldots,r$.  The gauge fields reside in real vector supermultiplets $V_a$,
and the gauge-invariant field-strengths are to be found in {\em twisted} chiral multiplets
$\Sigma_a = (\sigma_a, \lambda_{\pm,a}, \lambdab_{\pm,a}, D_a - i f_{01,a})$,
where $\sigma_a$ is a complex scalar, $\lambda_{\pm,a}$ are left/right-moving Weyl
fermions, $D_a$ is a real auxiliary field, and $f_{01,a}$ is the abelian gauge
field-strength.

We define the GLSM at a scale $\mu$ by a Lagrange density $\cL^\mu$ given
by a sum of two terms, the K\"ahler term $\cL_K^\mu$ and the 
{\em twisted superpotential} term $\cL_\Wt^\mu$. We take the K\"ahler term to be
\begin{equation}
\cL_K^\mu = \int d^4\theta\left( - \frac{1}{4} \sum_{i=1}^n \Phib^i
                        \exp\left(2 \sum_{a=1}^{r} Q^a_i V_a \right) \Phi^i
                      + \frac{1}{4 \mu^2 g(\mu)^2} \sum_{a=1}^{r} \Sigmab_a \Sigma_a\right),
\end{equation}
where $g(\mu)$ is the dimensionless coupling of the gauge theory. 
The tree-level twisted superpotential is given by
\begin{equation}
\cL_\Wt^\mu = \left[-\frac{i}{2\sqrt{2}}  \int d\tp d\tbm \sum_{a=1}^{r} \Sigma_a \tau^a(\mu)\right]  + \text{c.c.}.
\end{equation}
The $\tau^a(\mu) = i r^a(\mu) + \ff{\theta^a}{2\pi}$ are the parameters of the
model.  Each $\tau^a$ is a  combination of a Fayet-Iliopoulos (F-I) term $r^a$ and
a $\theta$-angle $\theta^a$. It is useful to define single-valued parameters 
$q_a = e^{2\pi i \tau^a}.$  For generic values of these parameters the moduli space of
classical vacua of the GLSM so defined is a toric variety.  The GLSM may be generalized
by including a superpotential $W(\Phi)$ which serves to restrict the moduli
space to a hypersurface or a complete intersection in the ambient toric variety.  
We will restrict attention to toric GLSMs, those with $W(\Phi) = 0$.

\subsection{Basic GLSM Properties}
                    
Let us begin with a brief review of the Higgs vacua of the GLSM.\footnote{There are no
photons and no Higgs mechanism in two dimensions, and a description based on Higgs vacua 
is only valid at weak coupling.  Fortunately, this is just where we will use it, and so 
we will ignore this subtlety in what follows.} This material is well known, and we refer 
the reader to \cite{W:phases,MP:summing} for further details.
The classical moduli space of a toric GLSM is obtained by solving the $D$-terms modulo the gauge
group as functions of the F-I parameters $r^a$.  One finds that there exists a cone 
$\K_c\subseteq \R^r$ where
the space of solutions to the $D$-terms is non-empty.\footnote{The D-term equations are 
$D^a = \sum_i Q^a_i |\phi^i|^2 - r^a,$ whence it follows that $\K_c$ is indeed a cone, the space
positively generated by the $n$ vectors $\Qv_i \in \R^r$.} For generic values of the $r^a \in \K_c$, 
the gauge group is completely broken, the $\sigma_a$ are massive, and the moduli space is a toric 
variety of complex dimension $d=n-r$.  The geometric properties of this toric variety vary 
smoothly with the $r^a$ away from co-dimension one sub-cones of $\overline{\K_c}$, where the gauge 
group is un-Higgsed and some or all of the $\sigma_a$ become massless.  These boundaries subdivide
$\K_c$ into a set of cones $\K_V$, indexed by a set of birationally equivalent toric varieties. 
There is a natural association between the varieties and the cones:  for $r^a \in \K_V$ the 
moduli space of the GLSM's classical vacua is the variety $V$. The cones $\K_V$ are termed {\em phases} 
of the GLSM.  By choosing the F-I parameters deep in the interior of any such phase, we obtain a weakly 
coupled theory whose low energy theory is that of a  NLSM with target-space $V$.  For reasons
that will become clear below, we will also refer to the complement of $\K_c$ as a phase.  

As the F-I terms are tuned to approach
a lower dimensional face of $\K_V$, the low energy description seems to break down as
$V$ becomes singular, or equivalently, there appear new massless degrees of freedom
corresponding to an un-Higgsed subgroup of the gauge group.  Quantum effects lift the
corresponding singularities when the un-Higgsed gauge group satisfies  $\sum_i Q_i \neq 0$, 
and even for gauge groups with $\sum_i Q_i=0$, the singularities are lifted for generic values 
of the corresponding $\theta$ angle.  Thus, all phases are smoothly connected, and the low-energy
NLSM description is smooth away from a complex co-dimension one subvariety in the space of
the $q_a$---the singular locus.  Of course, from the point of view of the GLSM there is no
real singularity on the singular locus.  However, we do expect that the theory is strongly
coupled on the singular locus, and strong coupling effects may invalidate results based on
the weakly coupled description.

In addition to the Higgs vacua, the GLSM possesses Coulomb vacua.  These are obtained when
some of the $\sigma_a$ acquire non-zero expectation values and give masses
to some or all of the matter fields. Integrating out these massive $\Phi^i$ multiplets 
leads to an effective interaction for the $\Sigma_a$ fields, which can be expressed 
in terms of an effective twisted superpotential $\Wte(\Sigma)$ \cite{W:phases,MP:summing}.
The solutions to $d \Wte(\sigma) =0$ are continuous if $\sum_i Q^a_i =0$ for all $\sigma_a$ 
with non-zero expectation values, and they are discrete otherwise.  The former exist only 
on the singular locus of the model, while the latter vary smoothly with the parameters.  
When all of the matter fields are massive, $\Wte(\Sigma)$ is given by\footnote{We have left off a conventional 
over-all factor of $-\frac{1}{4\pi\sqrt{2}}$. As far as our results are concerned, this
factor can be absorbed in the definition of the string coupling.}
\begin{equation}
\label{eq:wte}
\Wte =  \sum_{a=1}^{r} \Sigma_a
        \log\left[ \prod_{i=1}^n \left(\frac{1}{\exp(1) \mu}
                   \sum_{b=1}^{r} Q_i^b\Sigma_b\right)^{Q_i^a} /q_a\right].
\end{equation}
The vacua corresponding to $d\Wte = 0$ will occupy us for most of this note.  For future
reference, we give the equations of motion which follow from $d\Wte=0$:
\begin{equation}
\label{eq:teom}
\prod_{i|Q^a_i>0} \left(\frac{\xi_i}{\mu}\right)^{Q^a_i} = q_a \prod_{i|Q^a_i<0} \left(\frac{\xi_i}{\mu}\right)^{-Q^a_i},~~~a=1,\ldots,r,
\end{equation}
where $\xi_i = \sum_a Q^a_i\sigma_a$.  We will also have use for 
\begin{equation}
\label{eq:Hmatrix}
\cH^{ab} := \frac{\p^2 \Wte}{\p\sigma_a \p\sigma_b} = \sum_i \frac{Q^a_i Q^b_i}{\xi_i}.
\end{equation}
Of course, the Hessian of $\Wte$ is given by $H= \det \cH$.

The Coulomb vacua are derived by integrating out massive matter fields, and thus are only
reliable in the regions of the parameter space where these fields are indeed massive.
In principle, this may depend on the renormalization of the K\"ahler terms, but at least 
in the weak coupling regimes (i.e deep in the interior of some $\K_V$)  one may discern 
which Coulomb vacua are reliable.  At weak coupling, the $\phi$ mass term has the 
contribution $2 \sum_{i,a,b} |\phi^i|^2 Q^a_i Q^b_i \sigma_a\sigmab_b$, so that a critical 
point of $\Wte$ is not reliable if all the $\sigma_a$ are small.

When do these discrete Coulomb vacua arise?
By examining the equations of motion in eqn.(\ref{eq:teom}) it is clear that these are 
homogeneous in the $\sigma_a$ whenever $\sum_i Q^a_i = 0$ for all $a$, leading to a continuous
set of solutions for the $\sigma_a$.  We have not shown it, but it seems likely that if 
$\rank (Q) =r$ then this homogeneity is the only way to obtain a continuum of solutions. So,
we expect that $\Wte$ describes discrete Coulomb vacua whenever $\sum_i Q^a_i \neq 0$ for some $a$.
It is always possible to choose a basis for the action of the gauge group so that  $Q^a_i$ satisfy
$\sum_i Q^a_i = 0$ for $a >1$.  We will work in this basis, taking $\Delta = \sum_i Q^1_i$.
The condition for $\Wte$ to describe discrete Coulomb vacua is then just $\Delta \neq 0$.
When $\Delta =0$, the continuous solutions to $d\Wte =0$ emerge on the principal component
of the singular locus \cite{MP:summing}.

The $\Wte$ above describes the vacua where all of the $\Phi^i$ are massive.  Of course, 
there may also be {\em Coulomb-Higgs} vacua, where the gauge group is partially Higgsed.
Just as the Coulomb vacua described by $\Wte$, these may be labelled according to the 
space of $\sigma$ vevs as either {\em continuous} or {\em discrete}.  The former are found on 
various non-principal components of the singular locus of the model\cite{MP:summing}, while 
the latter, like their discrete Coulomb cousins, may be found in various phases.  We will not study 
the discrete Coulomb-Higgs vacua in this note.  However, when analyzing a particular phase one 
should be careful to check that the results are not invalidated by the presence of these vacua.

We will find it useful to characterize the phases of the GLSM by the types of vacua found at weak 
coupling.  Whenever a phase does not have any discrete Coulomb-Higgs vacua, we will 
refer to it as: 
\begin{itemize}
\item[-]  a {\em Geometric Phase} if its weak coupling limit has no reliable Coulomb vacua, and the vacua are
purely Higgs; 
\item[-]  a {\em Non-Geometric Phase} if the situation is reversed and there are no Higgs 
vacua; 
\item[-]  a {\em Mixed Phase} if both Coulomb and Higgs vacua are present at weak coupling.
\end{itemize}
These distinctions are important.  For example, in a model with a Non-Geometric Phase, 
eqn.(\ref{eq:corresult}) yields the correlators at any genus,  while in a model with a Geometric 
Phase, we may be able to compute the genus zero correlators by gauge instanton sums.  As we will
see below, in a Mixed Phase the correlators may be obtained by simply adding the Higgs and Coulomb 
contributions.  Note that a Non-Geometric Phase may only exist outside of $\K_c$, so it is only 
if $\K_c \not\simeq \R^r$, that our strongest results hold.  Happily, this holds for compact toric 
GLSMs.

%%%%%%%%%%%%%%%%%%%%%%%%%%%%%%%%%%%%%%%%%%%%%%%%%%%%%%%%%%%%%%%%
%%%%%%%%%%%%%%%%%%%%%%%%%%%%%%%%%%%%%%%%%%%%%%%%%%%%%%%%%%%%%%%%
\section{$A$-Model Localization on the Discrete Coulomb Branch} \label{s:proof}

\subsection{Twisting and Localization: Generalities}

The topological twisting of an $\N=(2,2)$ theory may be accomplished by shifting the 
spin connection on the world-sheet by the (ultraviolet) $R$-symmetry of the model.  
In effect, this produces a new theory by modifying the spins of the fields.  Let us now 
point out some basic consequences of this twisting in the context of the GLSM.  The twisted 
theory possesses a world-sheet anti-commuting scalar operator $\Qc$ which can be used to 
project the theory onto the $\Qc$-cohomology.  In the GLSM, this leads to topological 
observables parametrized by powers of the $\sigma_a$.  Another consequence of the twisting is 
that the path integral localizes onto the $\Qc$-invariant configurations---the SUSY vacua of 
the untwisted theory.  This property of {\em localization}\cite{W:topsig,W:mirtop,SZ:superlocal,BT:localization,CMR:ym} 
plays a crucial role in the study of TFT.  Roughly, this is the statement that the path 
integral will localize onto the vacua of the theory, and, under certain conditions, the 
contribution of a particular vacuum may be computed semi-classically.  This still leaves a 
difficult problem, especially when quantum vacua are involved. Of course, this is the case 
when we wish to study the correlators in a Mixed or Non-Geometric Phases of the GLSM.  
Fortunately, in that case the vacua are controlled by $\Wte(\Sigma)$, a quantity determined 
by holomorphy and 't Hooft anomaly matching.  

A further simplification makes the TFT computations tractable: since the singular locus 
is a complex co-dimension one variety in the space of the $q_a$, we expect that we should be 
able to compute correlators at weak coupling in {\em any} phase, and then unambiguously obtain 
the result for generic $q_a$ by analytic continuation.  Indeed,  this has been demonstrated for the 
Geometric Phases in \cite{MP:summing}.  As we will see below, the result also holds in more 
general situations involving the Coulomb vacua. 

\subsection{$A$-Twist Details}
On a Euclidean signature world-sheet $\Sigma_g$ the spin connection may be thought of as 
a $\GU(1)$ connection, and twisting amounts to shifting the Lorentz $\GU(1)$ charges of the fields by 
a combination of the $R$-charges.  In the toric GLSM the classical $\GU(1)_+ \times \GU(1)_-$ 
$R$-symmetry group leaves the superfields $\Phi^i$, $V_a$ invariant while acting
on the $\theta^\pm$ with charges $Q_\pm (\theta^\pm) = +1$, $Q_\mp (\theta^\pm) = 0$.
All other charges are determined by this choice, and, in particular, 
$Q_\pm(\sigma_a) = \pm 1$.  When $\Delta\neq 0$, this classical $R$-symmetry suffers from 
an anomaly in the presence of gauge instantons.  That means that only the vector combination may be
used to obtain a consistent twisted theory.  This is the $A$-model, obtained by
twisting with $Q_V = \ff{1}{2} (Q_++Q_-)$.  If we designate the Lorentz charges of the 
fields by $Q_L$,  the new Lorentz charges are $Q_L' = Q_L - Q_V$.  Applying this to the 
fields of the GLSM we find that the $\phi^i$, $\sigma_a$ remain world-sheet scalars, 
but the spins of the fermions are shifted. The $\psi_+,\psib_-,\lambda_-,\lambdab_+$ become world-sheet
one-forms, while $\psi_-,\psib_+,\lambda_+,\lambdab_-$ become world-sheet scalars \cite{W:mirtop}:
\begin{equation}
 \begin{array}{cccccc}
  \psi_+  & \to & \psi_z      & \lambda_+  & \to & \eta \\
  \psi_-  & \to & \chi        & \lambda_-  & \to & \rho_{\zb} \\
  \psib_+ & \to & \chib       & \lambdab_+ & \to & \rhob_{z} \\
  \psib_- & \to & \psib_{\zb} & \lambdab_- & \to & \etab.
 \end{array}
\end{equation}

\subsection{Localization}
We are interested in computing expectation values of the form $\la F(\sigma)\ra_g$ in
the twisted theory.  As described above, to perform these computations we deform the 
model to weak coupling, where we have a reasonable handle on the vacua of the theory.
Since the TFT path integral localizes onto the vacua, we may compute the correlators 
by summing contributions from the vacua.  Let us suppose we are working in a phase of 
the GLSM without discrete Coulomb-Higgs vacua.  At weak coupling, the path integral 
then receives contributions from the Higgs vacua---the gauge instantons, and the 
discrete Coulomb vacua:
\begin{equation}
\label{eq:PI}
Z = Z^{\text{Higgs}} + Z^{\text{Coulomb}}.
\end{equation}
At $g=0$, the contribution from the gauge instantons may be determined by the methods 
of Morrison and Plesser \cite{MP:summing}, and we will now show how to compute 
$Z^{\text{Coulomb}}$ at any genus.

Since we wish to compute $\la F(\sigma)\ra_g$, we may first perform the integration over 
the $\Phi^i$ multiplets.  As we have argued above, in the untwisted theory this leads 
to a factor of $\exp\left(-\int d^2 z \cL_{\Wte}\right)$ in the path integral over 
the $\Sigma_a$ multiplets.  When we perform this integration in the twisted theory, we 
find a similar result, but we must be careful of one subtlety:  the {\em zero modes} 
of the $\Phi^i$ multiplets.  In the untwisted theory the $\psi_\pm,\psib_\pm$ had no zero 
modes, and, aside from the factor of $\exp\left(-\int d^2 z\cL_{\Wte}\right)$ and a 
deformation of the (irrelevant) K\"ahler term of the $\Sigma_a$, the one-loop fermion 
determinants cancelled the bosonic ones as a consequence of the $\N=(2,2)$ SUSY.  In the 
twisted theory, while the non-zero modes of the $\Phi^i$ multiplets continue to be paired up 
just as in the untwisted theory, the zero modes of the fermions no longer pair up with 
the $\phi^i$ zero modes.  To deal with this subtlety, we will separate out the integral over
the $\Phi^i$ zero modes.  Of course, integrating out the non-zero modes will still lead to
the factor of $\exp\left(-\int d^2 z \cL_{\Wte}\right)$.

We are now in a position to apply the standard localization arguments to the Coulomb vacua.
The corresponding $\Qc$-invariant configurations are a subset of the field configurations
\begin{equation}
 \phi^i =0, ~~~ f_a = 0, ~~~ \p_z \sigma_a = \p_{\zb} \sigma_a = 0, ~~~ d \Wte = 0.
\end{equation}
A solution to $d \Wte(\sigma)=0$ does not necessarily correspond to a Coulomb vacuum, and
at generic $q_a$ it is difficult to determine which of the solutions to $d \Wte(\sigma) = 0$ 
are reliable.  However, at weak coupling, i.e. deep in the interior of a phase, we may answer 
this question unambiguously: a $\sigma$ vacuum is trustworthy only if in the weak coupling 
limit the corresponding $|\sigma_a|$ grow in such a way that all of the $\Phi^i$ multiplets 
may be consistently integrated out.  We will label these reliable vacua by $\sigmah$.  The 
contribution to the path integral is then
\begin{equation}
Z^{\text{Coulomb}} = \sum_{\sigmah} Z(\sigmah),
\end{equation}
and we can compute $Z(\sigmah)$ at weak coupling by expanding in fluctuations about 
the $\sigmah$ vacuum.  Integration over the massive modes of the $\Sigma$ multiplets leads 
to determinants that exactly cancel between the bosons and fermions (this familiar fact may 
be traced back to the primordial $\N=(2,2)$ SUSY), and we are left with an integral over the 
zero-modes, which factorizes into an integral over the $\Phi$ fluctuations and an 
integral over $\Sigma$ fluctuations:
\begin{equation}
Z(\sigmah) = \int [D\Phi] \int [D\Sigma] \exp(-S_\Phi -S_\Sigma) = \int [D\Phi] \exp(-S_\Phi) \int[D\Sigma] \exp(-S_\Sigma) 
           = Z_\Phi(\sigmah) Z_\Sigma(\sigmah).
\end{equation}
The terms in the action are given by
\begin{eqnarray}
-S_\Phi & = &  \sum_i \left\{ V_\Sigma \left[  -2 |\xi_i|^2 |\phi^i|^2 + \sqrt{2} \chi^i \xib_i \chib^i\right] 
               + \sqrt{2} \int_{\Sigma_g} \psi^i_z \xi_i \wedge \psib^i_{\zb} \right\},\nonumber\\
-S_\Sigma & = & \sum_{a,b} \left\{ V_\Sigma \left[ -4\mu^2\sigmab_a \left(\cH^\dag\cH\right)^{ab} \sigma_b 
                                                 + \etab_a 2 \cHb^{ab} \eta_b\right]
                                                 + \int_{\Sigma_g} \rhob_{z,a} 2\cH^{ab} \wedge \rho_{\zb,b} \right\},
\end{eqnarray}
where $V_\Sigma$ is the volume of $\Sigma_g$, and $\xi_i$ and $\cH^{ab}$---the latter defined in eqn.(\ref{eq:Hmatrix})---
are to be evaluated at $\sigma = \sigmah$.  We can now evaluate the Gaussian integrals.  For $Z_\Phi$ we find
\begin{equation}
Z_\Phi = Z_\phi Z_\chi Z_\psi,
\end{equation}
\begin{equation}
 Z_\phi^{-1} =  V_\Sigma^n \prod_i \left|\xi_i\right|^2, ~~~
 Z_\chi =  V_\Sigma^n \prod_i \xib_i, ~~~
 Z_\psi =  \prod_i \xi_i^{g}.
\end{equation}
We have counted the $1$ zero mode of $\chi$ (there is just one constant function on a compact $\Sigma_g$)
and the $g$ zero modes of the $\psi^i_z$ (there are $g$ holomorphic one-forms on $\Sigma_g$).  Similarly, 
\begin{equation}
Z_\Sigma = Z_\sigma Z_\eta Z_\rho,
\end{equation}
with
\begin{equation}
Z_\sigma^{-1}  =   V_\Sigma^r \left|H\right|^2 ,~~~ 
Z_\eta   =  V_\Sigma^r \Hb,~~~ 
Z_\rho   =  H^{g},
\end{equation}
where $H = \det \cH$.
Putting it all together, we find the measure in eqn.(\ref{eq:corresult}). 

The careful reader will note that we have ignored any subtleties associated to gauge invariance.
This simplification follows because the computations are performed on the Coulomb 
branch, where the condition $f_a = 0$ and the Riemann-Roch theorem ensure that our computation of 
$Z_\Phi$ is correct.  We have also neglected various constants which may be absorbed into an overall 
normalization of the correlators or a re-definition of the string coupling constant.  Furthermore, 
the sign of the fermion integration measure has been chosen to match results from the Higgs Phase 
computations at genus zero.

In what follows, we will work in units of the scale $\mu$.  This scale plays an important role
in the untwisted theory, but upon twisting it becomes superfluous, essentially because the TFT is a 
theory at zero energy.  The scale can be important if one wants to make connections with the untwisted 
theory, in which case it is easy to restore in our formulas.

%%%%%%%%%%%%%%%%%%%%%%%%%%%%%%%%%%%%%%%%%%%%%%%%%%%%%%%%%%%%%%%
%%%%%%%%%%%%%%%%%%%%%%%%%%%%%%%%%%%%%%%%%%%%%%%%%%%%%%%%%%%%%%%
\section{A Few Applications to Compact Toric GLSMs} \label{s:examples}

As discussed above, we expect that whenever $\K_c \not\simeq \R^r$, eqn.(\ref{eq:corresult})
directly gives the correlators at arbitrary genus.  This makes it useful for elucidating 
various properties of the compact toric GLSM $A$-model correlators, as well as actual 
computations.  We illustrate this in this section.

\subsection{Some Properties of the Correlators}
We can easily demonstrate some important properties of these GLSM $A$-model correlators
from the explicit form.  Perhaps the simplest observation is that the result presents 
the correlators as a sum over all the solutions to a system of polynomial equations with finitely
many common zeroes.  This finite sum has a natural expansion in terms of symmetric functions,  and, thus,
it  is clear that the correlators are meromorphic functions of the $q_a$.  This is not obvious 
from the form of the instanton sum in a Geometric Phase.  Another equally simple but
important observation is that the quantum cohomology relations, which are just the equations
of motion in eqn.(\ref{eq:teom}) considered as operator relations, obviously hold.  This 
should also be compared with the Geometric phase computation, where this is a non-trivial 
combinatorics result\cite{MP:summing,ME:glsmtach}.  Below, we give a few more technical observations. 

\subsubsection{TFT Factorization}
The $A$-twist reduces the Hilbert space of the GLSM to a vector space of dimension $N_v$, where $N_v$ is
the number of discrete Coulomb vacua.  The operators $\sigma_a$ are now simply $N_v\times N_v$ matrices,
and correlators are obtained by taking a matrix trace:
\begin{equation}
\la F(\sigma) \ra_g = \Tr \left[ F(\sigma) \left(\Ht(\sigma)\right)^{g-1}\right].
\end{equation}
These correlators are easily shown to satisfy the factorization axioms of topological field theory.
These axioms state that if we choose a complete basis of states $|i\ra$, and the corresponding 
operators $\phi_i$ have the metric $\eta_{ij} = \la \phi_i \phi_j \ra_0,$ with inverse $\eta^{ij}$,
then
\begin{enumerate}
 \item if $F(\sigma) = f_1(\sigma) f_2(\sigma)$, then
       \begin{equation}
       \la F(\sigma) \ra_g = \sum_{ij} \la f_1(\sigma) \phi_i \ra_{g'} \eta^{ij} \la \phi_j f_2(\sigma) \ra_{g-g'},
       \end{equation}
 and 
 \item for any $F(\sigma)$ 
       \begin{equation}
       \la F(\sigma) \ra_g = \sum_{ij} \eta^{ij} \la \phi_i \phi_j F(\sigma) \ra_{g-1}.
       \end{equation}
\end{enumerate} 
These properties are apparent in a basis of states corresponding to the $N_v$ $\sigma$-vacua.  The
state operator correspondence is $|i\ra \leftrightarrow \phi_i = \delta_{\sigma,\sigma_i}$, 
where $\sigma_i$ is the value of $\sigma$ in the $i$-th vacuum.  In this 
basis the operator $\sigma$ is diagonal, and $\eta^{ij} = \Ht(\sigma_i) \delta^{ij}$. 
Factorization follows immediately.  Thus, as expected, any genus correlator may be obtained from 
the $g=0$ results.

\subsubsection{The Ghost Number Selection Rule}
These $A$-model correlators obey a simple selection rule.  Working in the basis where 
\begin{equation}
\sum_i Q^1_i = \Delta ~~\text{and}~~ \sum_i Q^a_i =0 ~~\text{for}~~ a>1, 
\end{equation}
we may write the $\sigma$ equations of motion in
the form $\sigma_a = \omega_a \sigma_1$ for $a>2$, where $\omega_a$ are now determined by solving $r-1$
polynomial equations, and $\sigma_1$ satisfies $\sigma_1^\Delta = q_1 s(\omega)$ for some $s(\omega)$.
Thus, the sum over the vacua includes a sum over the $\Delta$-th roots of unity.  Since $\Ht$ has
degree $d=n-r$, if $F(\sigma)$ has degree $s$, then $\la F(\sigma)\ra_g$ is non-zero only 
if $s + d (g-1) = m \Delta$ for some integer $m$, in which case $\la F(\sigma)\ra_g \sim q_1^m.$  
This selection rule is just the ghost number selection rule familiar from TFTs in general and GLSMs in 
particular.

\subsubsection{The All Genus Correlation Function}
Although factorization makes this exercise  purely one of convenience, we can easily sum over the 
genera to obtain
\begin{equation}
 \la F(\sigma) \ra = \sum_{g\ge 0} \lambda^{2g-2} \la F(\sigma)\ra_g = \Tr\left[ \frac{F(\sigma)}{\lambda^2 \Ht} \frac{1}{1-\lambda^2\Ht}\right].
\end{equation}
From the selection rule above it follows that if 
$F(\sigma)$ has degree $s$ then 
\begin{equation}
 \la F(\sigma) \ra = q_1^{s/\Delta} f\left( \left(q_1^d\lambda^{2\Delta}\right)^{1/gcd(d,\Delta)}\right).
\end{equation}

\subsubsection{The Quantum Restriction Formula}
Given a Calabi-Yau hypersurface in a toric variety $V$, there exists a simple method for obtaining
the ``toric'' subset \cite{AGM} of the $g=0$  $A$-model correlators for the Calabi-Yau model.
These correlators can be computed by the {\em quantum restriction formula} of \cite{MP:summing},
which expresses a hypersurface toric $A$-model correlator, denoted by $\lad F(\sigma)\rad$, to a sum 
over the $A$-model correlators for $V$:
\begin{equation}
\lad F(\sigma) \rad_{g=0} = \la F(\sigma) \frac{-K}{1-K}\ra_{g=0},
\end{equation}
where $-K$ is the operator corresponding to the anti-canonical divisor on $V$, given by $-K = \sum_i \xi_i$.
Using our form of the correlators on $V$, it follows that
\begin{equation}
\lad F(\sigma) \rad_{g=0} = \Tr \left[ \frac{ F(\sigma)}{\Ht} \frac{-K}{1-K}\right].
\end{equation}

\subsection{Two Examples}

In this section we will apply our simple result and the observations above to 
two examples.  These models are not difficult to solve, but they illustrate
some techniques and ideas that should be useful even in much more intricate examples.

\subsubsection{$A$-model correlators for $\P^4$.}
Let us start with the canonical GLSM example:  $\P^4$.  This is a one
parameter model with $ Q = (1,1,1,1,1).$  The equation of motion from eqn.(\ref{eq:teom}) is
just $\sigma^5 =  q,$  and $\Ht = 5 \sigma^4,$ yielding
\begin{equation}
\label{eq:p4cor}
\la \sigma^a \ra_g = 5^{g-1}\sum_{\sigma|\sigma^5 = q} \sigma^{a+4(g-1)}.
\end{equation}
The correlators satisfy the selection rule discussed earlier:
\begin{equation*}
\la \sigma^a\ra_g = 0~~\text{unless}~~a+4(g-1) = 5 n~~\text{for some integer}~n,
\end{equation*}
in which case
\begin{equation}
\la \sigma^{5n+4(1-g)}\ra_g = 5^g q^n.
\end{equation}

The all-genus correlation function is given by
\begin{equation}
\la \sigma^a \ra = \sum_{g\ge 0} \lambda^{2 g-2} \la \sigma^a \ra_g.
\end{equation}
Evaluating this for $ 0\le a \le 4$, we find
\begin{eqnarray}
\la \sigma^a\ra & = & \frac{5 (5 q \lambda^2)^a }{1-5^5q^4\lambda^{10}}, a=0,\ldots, 3,\nonumber\\
\la \sigma^4\ra & = & \frac{\lambda^{-2}}{1- 5^5q^4\lambda^{10}}.
\end{eqnarray}
The intriguing pole at $q^4\lambda^{10} = 5^{-5}$ agrees with the findings of \cite{W:tphase}.
The interpretation of this pole is far from clear.  While we might expect such a pole in a 
topological string theory, where it could be a manifestation of non-perturbative effects in $\lambda$,
we have not coupled the model to $d=2$ gravity, and thus any string-based interpretation does not seem
appropriate.

Finally, we can use the quantum restriction formula to compute the unique $A$-model correlator on the
quintic in $\P^4$.  The anti-canonical divisor corresponds to $-K = 5 \sigma$, and we find
\begin{equation}
\lad \sigma^3 \rad_{g=0} = \Tr \frac{\sigma^3}{5\sigma^4} \frac{5\sigma}{1+5\sigma} = \Tr\frac{1}{1+(5\sigma)^5} = \frac{5}{1+5^5q}.
\end{equation}

\subsubsection{A Two Parameter Example}
This is another example that has been studied in detail in \cite{MP:summing}.  This GLSM corresponds
to the toric variety obtained by resolving the curve of $\Z_2$ singularities in the weighted projective
space $\P^4_{1,1,2,2,2}$.  The GLSM has $n=6$, $r=2$ and charges
\begin{equation}
Q = \left(\begin{array}{cccccc} 0 & 0 & 1 & 1 & 1&  1 \\
                                1 & 1 & 0 & 0 & 0 &-2
          \end{array}
    \right).
\end{equation}
Obviously, $\Delta = 4$, and  $\Ht = 8 \sigma_1^3\sigma_2$.
Letting $\sigma_2 = \omega \sigma_1$, the equations of motion $d\Wte=0$ may be written as
\begin{equation}
\sigma_1^4 = \frac{q_1}{1-2\omega},
\end{equation}
and 
\begin{equation}
P(\omega) = \omega^2-q_2(1-2\omega)^2 = 0.
\end{equation}
The selection rule implies that $\la \sigma_1^a \sigma_2^b\ra_g$ is zero unless $a+b = 4(m+1)$,
and if $m\ge 0$, we have
\begin{equation}
\la \sigma_1^{4(m+1)-b} \sigma_2^b \ra = \sum_{g\ge 0} \Tr \left[ \sigma_1^{4(m+g)}\omega^{b+g-1} (8\lambda^2)^{g-1}\right],
\end{equation}
which we can reduce to a trace on the roots of $P(\omega)$, denoted by $\Tr'$:
\begin{equation}
\la \sigma_1^{4(m+1)-b}\sigma_2^b\ra = \frac{ q_1^m}{2 \lambda^2(1- (8\lambda^2q_1)^2 q_2 )} 
              \Tr'\left[\frac{\omega^{b-1}(1+(8\lambda^2q_1-2)\omega)}{(1-2\omega)^{m+1}}\right].
\end{equation}
Again, we observe the interesting $\lambda$-dependent pole.

At genus zero the above expression simplifies to 
\begin{equation}
\la \sigma_1^{4(m+1)-b}\sigma_2^b\ra_{g=0} = \frac{q_1^m}{2} \Tr'\frac{\omega^{b-1}}{(1-2\omega)^m}.
\end{equation}
We can again use the quantum restriction formula to compute correlators on the anti-canonical hypersurface.
We have $-K = 4\sigma_1$, and
\begin{equation}
\lad \sigma_1^{3-j}\sigma_2^j\rad_{g=0} = 4 \la \frac{\sigma^{4-j}\sigma_2^j}{1 + (4\sigma_1)^4} \ra_{g=0} 
                                        = 2 \Tr' \frac{\omega^{j-1}(1-2\omega)}{1-4^4q_1 -2\omega}. 
\end{equation}

In this and other two-parameter models it is convenient to rewrite the $\Tr'$ as a contour
integral:
\begin{equation}
\Tr' f(\omega) = \sum_{\omegah|P(\omegah) =0} \oint_{C(\omegah)} \frac{d\omega}{2\pi i} \frac{f(\omega) P'(\omega)}{P(\omega)},
\end{equation}
where $C(\omegah)$ is a small contour about $\omega=\omegah$.  This form makes it easy to evaluate the
traces.  In the case of more than two parameters, more sophisticated residue techniques may be 
applied \cite{VAFA:tlg,GH:alggeom}.
Applying this to the case at hand, 
\begin{equation}
\lad \sigma_1^{3-j}\sigma_2^j\rad_{g=0} = \sum_{\omegah|P(\omegah) =0} \oint_{C(\omegah)} \frac{d\omega}{2\pi i}
                                        \frac{4 \omega^{j}}{(1-4^4q_1 -2\omega)P(\omega)}. 
\end{equation}
Pulling the contour off the roots of $P(\omega)$, the correlators are written as
\begin{equation}
\lad \sigma_1^{3-j}\sigma_2^j\rad_{g=0} = 2 \left. \frac{\omega^j}{P(\omega)}\right|_{\omega=\frac{1-4^4q_1}{2}}
               +2 \Res\left.\left\{ \frac{\omega^j}{(\omega-\frac{1-4^4q_1}{2})P(\omega)}\right\}\right|_{\omega=\infty}.
\end{equation}
Straightforward algebra yields
\begin{eqnarray}
\lad \sigma_1^{3}\rad_{g=0}  & = &  \frac{8}{D},\nonumber\\
\lad \sigma_1^{2}\sigma_2\rad_{g=0}  & = & \frac{4 (1-2^8q_1)}{D},\nonumber\\
\lad \sigma_1 \sigma_2^2\rad_{g=0}  & = &  \frac{8q_2(2^9q_1 -1)}{(1-4q_2)D},\nonumber\\
\lad \sigma_2^3\rad_{g=0}  & = &  \frac{4q_2(1+4q_2-2^8q_1-3072 q_1 q_2)}{(1-4q_2)^2D},
\end{eqnarray}
and $D = (1-2^8 q_1)^2 - 2^{18} q_1^2 q_2$.
This reproduces the results of eqn.(4.28) of \cite{MP:summing}.\footnote{Our expression corrects
a sign error in the last correlator in eqn.(4.28) of \cite{MP:summing}. }

%%%%%%%%%%%%%%%%%%%%%%%%%%%%%%%%%%%%%%%%%%%%%%%%%%%%%%%%%%%%%%%%
%%%%%%%%%%%%%%%%%%%%%%%%%%%%%%%%%%%%%%%%%%%%%%%%%%%%%%%%%%%%%%%%
\section{A Non-Compact Example} \label{s:zforreal}

Having examined the properties of models with a Non-Geometric Phase, we now turn to models
where Higgs vacua are present in every phase.  Sadly, this means that with the
current technology we will need to restrict attention to genus zero correlators, but, 
nevertheless, we will be able to uncover some surprises.

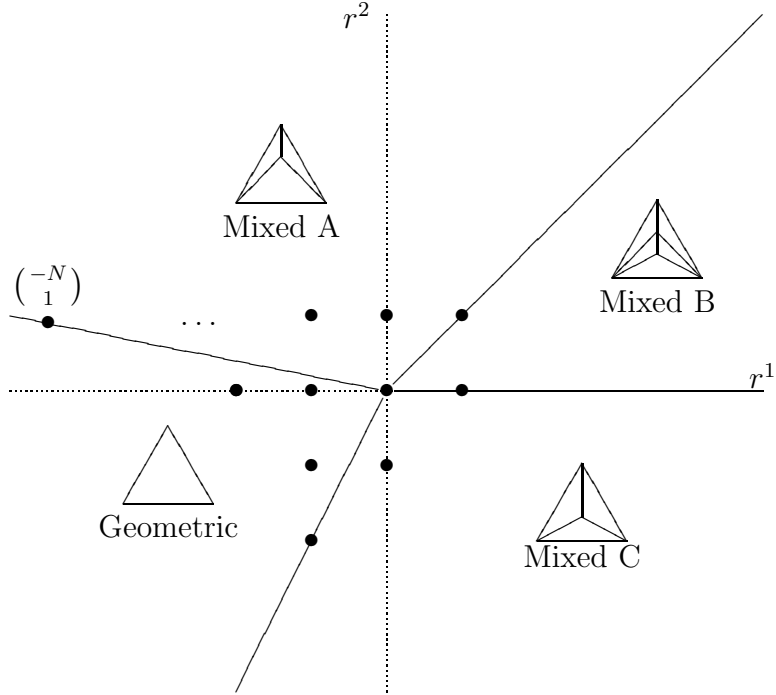
\begin{figure}
\[
\begin{xy} <1.0mm,0mm>:
  (0,0)*{\bullet} ="o", (50,0)*{}="1", (50,50)*{}="2", (-50,10)*{}="3",
  (-20,-40)*{}="4", (-50,0)*{}="7", (10,0)*{\bullet}, (0,10)*{\bullet},
  (-10,-20)*{\bullet}, (10,10)*{\bullet}, (-10,10)*{\bullet},
  (-20,0)*{\bullet}, (-20,0)*{\bullet}, (-10,0)*{\bullet},
  (0,-10)*{\bullet}, (-10,-10)*{\bullet}, (-45, 9)*{\bullet},
  (0,50)*{}="5", (0,-40)*{}="6", (50,2)*{r^1}, (-4,50)*{r^2}, (-25,9)*{\ldots},
  (-45,14)*{\left(\genfrac{}{}{0pt}{}{-N}{1}\right)}, 
  (30,15)*\xybox{ <0.3mm,0mm>:
  (0,0)*{}="1",(40,0)*{}="2",(20,34.64)*{}="3", (20,10.66)*{}="4", (20,20.43)*{}="5",
  \ar@{-}|{} "1";"2" \ar@{-}|{} "2";"3"
  \ar@{-}|{} "3";"1" \ar@{-}|{} "3";"5"
  \ar@{-}|{} "5";"4" \ar@{-}|{} "4";"1"
  \ar@{-}|{} "4";"2" \ar@{-}|{} "5";"1"
  \ar@{-}|{} "5";"2" },
  (36,12)*{\text{Mixed B}},
  (-20,25)*\xybox{ <0.3mm,0mm>:
  (0,0)*{}="1",(40,0)*{}="2",(20,34.64)*{}="3", (20,20.43)*{}="5",
  \ar@{-}|{} "1";"2" \ar@{-}|{} "2";"3"
  \ar@{-}|{} "3";"1" \ar@{-}|{} "5";"1"
  \ar@{-}|{} "5";"2" \ar@{-}|{} "5";"3"
  },
  (-14,22)*{\text{Mixed A}},
  (20,-20)*\xybox{ <0.3mm,0mm>:
  (0,0)*{}="1",(40,0)*{}="2",(20,34.64)*{}="3", (20,10.66)*{}="4",
  \ar@{-}|{} "1";"2" \ar@{-}|{} "2";"3"
  \ar@{-}|{} "3";"1" \ar@{-}|{} "4";"1"
  \ar@{-}|{} "4";"2" \ar@{-}|{} "4";"3"
  },
  (26,-22)*{\text{Mixed C}},
  (-35,-15)*\xybox{ <0.3mm,0mm>:
  (0,0)*{}="1",(40,0)*{}="2",(20,34.64)*{}="3"
  \ar@{-}|{} "1";"2" \ar@{-}|{} "2";"3"
  \ar@{-}|{} "3";"1"
  },
  (-29,-18)*{\text{Geometric}}
\ar@{-}|{} "o"; "1" \ar@{-}|{} "o"; "2"
\ar@{-}|{} "o"; "3" \ar@{-}|{} "o"; "4"
\ar@{.}|{} "5"; "6" \ar@{.}|{} "7"; "1"
\end{xy}
\]
\caption{Phases of the GLSM.}
\label{fig:phases}
\end{figure}

We will work with the example studied at length in \cite{ME:glsmtach}.
This GLSM has $n=5$, $r=2$ and charges
\begin{equation}
 Q = \left(\begin{array}{ccccc}
                 1 & 1 & 1 &  -N & -1 \\
                 0 & 0 & 1 &   1 & -2
               \end{array}
         \right).
\end{equation}
There are four classical phases, each corresponding to a triangulation
of a toric fan.  The fan without any subdivisions is an orbifold, $\Orbt$,  the
partially subdivided fans correspond to partial resolutions, and the completely
subdivided fan is the smooth phase.  These phases are depicted in Fig. (\ref{fig:phases}).
We will assume $N>2$, and we have labelled the phases according to the 
Geometric-Mixed terminology defined above.  This model has a continuous Coulomb
branch which emerges for small $|q_1|$ and $q_2 = 1/4$,  and, naively, one would
expect that some observables in the TFT will be sensitive to this singularity.

The $A$-model correlators of interest are the $Y_{a,b} = \la \sigma_1^a \sigma_2^b\ra_{g=0}$.
The ghost number selection rule requires that for a non-zero correlator $a+b= 3 + (2-N)n$. 
In \cite{ME:glsmtach}, we were able to compute the $Y_{3+(2-N) n-b,b}$ correlators for $n<0$
by summing the instantons in the Geometric Phase, and we found that these correlators could be
put into a ``Coulomb'' form:\footnote{We call this the ``Coulomb'' form because it is the answer
that one would get by naive application of eqn.(\ref{eq:corresult}).}
\begin{equation}
\label{eq:yhigs}
Y_{3+(2-N)n-b,b}^{\text{Geometric}} = q_1^n \Tr' \frac{ \omega^b s(\omega)^n}{3N+1 + 2N\omega},
\end{equation}
where $\Tr'$ is to be taken over the roots of 
\begin{equation}
\label{eq:P}
P(\omega) = (1+\omega)(-N+\omega) - q_2 (1+2\omega)^2,
\end{equation}
and $s(\omega)$  is given by
\begin{equation}
\label{eq:s}
s(\omega) = \frac{ (-1-2\omega)(-N+\omega)^N}{1+\omega}.
\end{equation}
For later convenience, we will re-write these as a contour integral:
\begin{equation}
\label{eq:firsty}
Y_{3+(2-N)n -b, b} =  \sum_{\omegah=\omega_+,\omega_-} \oint_{C(\omegah)} \frac{d\omega}{2\pi i}
                      \frac{\omega^b s(\omega)^n}{(1+2\omega)P(\omega)},
\end{equation}
where $C(\omegah)$ is a small contour about $\omega =\omegah$, and $\omega_\pm$ are the
roots of $P(\omega)$.  

The computation of the $n = 0$ correlators is complicated by the non-compactness of
the orbifold, and in \cite{ME:glsmtach} we circumvented that problem by using the
quantum cohomology relations to determine the $Y_{3-b,b}$.  As we saw above, these
relations are powerful, and it is easy to show \cite{ME:glsmtach} that {\em if}
the $Y_{3-b,b}$ are determined from the $Y_{3+(2-N)n-b,b}$ by the relations, they must be of
given by eqn.(\ref{eq:yhigs}) with $n=0$.  Upon computing the trace, one finds that the 
$Y_{3-b,b}$ so determined are sensitive to the $q_2 = 1/4$ singularity.  As we will show below, 
our basic assumption was incorrect. The quantum cohomology relations simply do not hold!  

Even without further computations, there are several reasons to suspect the validity of this result.  
First, the $Y_{3-b,b}$ are independent of  $q_1$ and thus, if they are sensitive to the $q_2 = 1/4$
singularity at small $|q_1|$, they are equally singular at $q_2 = 1/4$ for {\em arbitrary} $q_1$. 
However, we are hard pressed to explain the singularity at large $|q_1|$ and $q_2=1/4$.  After all,
this is deep in the weakly coupled regime of the Geometric Phase, where a classical analysis is
reliable and does not reveal any singularities.  In addition, we know that the gauge instantons are 
labelled by sets of integers $\nv \in \left(\R^d\right)^\vee$, and each such instanton contributes a term 
$Y^{\nv} \prod_a q_a^{n_a} $  to the correlator.  In a particular phase $\K_V$, the 
instanton numbers corresponding to non-zero $Y^{\nv}$ must lie in the 
{\em dual cone} defined by
\begin{equation}
 \K_V^\vee = \left\{ \nv \in \left(\R^d\right)^\vee | \la \nv, \rv \ra \ge 0~~~\forall~~~ \rv \in \K_V\right\}.
\end{equation}
It is easy to see that in the Geometric Phase the only instantons that can contribute to $Y_{3-b,b}$ have $\nv=0$.  
Hence, one would expect $Y_{3-b,b}$ to be constants, and any $q_2$ dependence, let alone a singular one, is strange 
indeed.  This would seem to indicate that the quantum cohomology relations are violated whenever the $Y_{3-b,b}$ 
correlators are involved. 

To explore this further, let us now work out the correlators in one of the Mixed Phases.  
We will choose the phase $A$, but the computation may be easily repeated for other phases. 
First, let us compute the contribution from the Coulomb vacua.
For this model, $\Ht$ 
is given by
\begin{equation}
\Ht = (N-2) \sigma_1^2((3N+1)\sigma_1 + 2N\sigma_2),
\end{equation}
and the equations of motion that follow from $d\Wte = 0$ are
\begin{eqnarray}
\left(\sigma_1 + \sigma_2\right) \left(\sigma_2 -N \sigma_1\right)  & = &
  q_2 \left(\sigma_1 + 2 \sigma_2\right)^2 , \nonumber\\
\sigma_1^2\left(\sigma_1 + \sigma_2\right) & = &
 -q_1 \left(\sigma_2 -N\sigma_1\right)^N \left(\sigma_1 + 2\sigma_2\right).
\end{eqnarray}
As in the previous section, we may parametrize the solutions by $\sigma_1$ and 
the ratio $\omega = \sigma_2/\sigma_1$:
\begin{eqnarray}    
P(\omega_\pm) & = & 0
                 \nonumber\\
\sigma_{1,\pm;p} & = & \zeta^p \left(q_1 s(\omega_\pm)\right)^{\frac{1}{2-N}}, \nonumber\\
\sigma_{2,\pm;p} & = & \omega_\pm \sigma_{1,\pm;p},
\end{eqnarray}
where $\zeta = e^{\frac{2 \pi i}{ N-2}}$,  $p=0,\ldots, N-1$, and $P(\omega)$ and $s(\omega)$ are
as in eqns.(\ref{eq:P},\ref{eq:s}).

Let us consider the solutions $(\sigma_{1,\pm;p},\sigma_{2,\pm;p})$ in the weak coupling limit.
Weak coupling in the Mixed Phase A corresponds to 
\begin{equation}
 |q_2|^{-N} \gg |q_1| \gg |q_2|,
\end{equation}
and in particular, $|q_2| \to 0$.  In this limit the $\sigma$-vacua have a simple structure:
\begin{eqnarray}
\label{eq:Proots}
\omega_+ & \to & -1 -\frac{1}{N+1}q_2, \nonumber\\
\omega_- & \to & N +\frac{(2N+1)^2}{N+1}q_2,
\end{eqnarray}
and hence
\begin{eqnarray}
\sigma_{1;+}^{2-N}  & \to &  q_1 q_2^{-1} (-N-1)^{N+1}, \nonumber\\
\sigma_{2;-}^{2-N}  & \to & -q_1 q_2^N \frac{ (2 N+1)^{2N+1}}{(N+1)^{N+1}}.
\end{eqnarray}
Thus, we see that in the weak coupling limit of the Mixed phase the $N-2$  ``$-$'' critical points
of $\Wte$  have growing $\sigma$ vevs, while the $N-2$ ``$+$'' critical points have decreasing 
$\sigma$ vevs.  Thus, only the ``$-$'' solutions correspond to actual Coulomb vacua, and their
contribution, $Y_{3+(2-N)n-b,b}^{\text{Coulomb}}$,  is given by the $\omega_-$ contribution in
eqn.(\ref{eq:firsty}): 
\begin{equation}
Y_{3+(2-N) n -b, b}^{\text{Coulomb}} = q_1^n \oint_{C(\omega_-)} \frac{\omega^b s(\omega)^n }{(1+2\omega)P(\omega)}.
\end{equation}

Next, we consider the Higgs contribution. Unlike the Coulomb computation, which gives the same form 
regardless of whether $n=0$ or $n<0$, here this distinction makes a crucial difference.
First, let us consider the situation when $n<0$.  Using the standard toric techniques of
Morrison and Plesser, we can perform the instanton sum and evaluate the Higgs branch contribution
to the correlators.  Performing the requisite toric intersection computations, we reduce the correlators
to a single sum:
\begin{equation}
Y_{3+(2-N) n -b, b}^{\text{Higgs}} = - q_1^n \sum_{m=-n}^{\infty} \oint_{C(-1)} \frac{d\omega}{2\pi i}
          \frac{\omega^b s(\omega)^n R^m}{(1+\omega)(-N+\omega)(-1-2\omega)},
\end{equation}
where 
\begin{equation}
R = q_2 \frac{(1+2\omega)^2}{(1+\omega)(-N+\omega)},
\end{equation}
and $C(-1)$ is a small contour about 
$\omega=-1$:  $\omega = -1 + \ep e^{i\theta}$.  For uniform convergence we must have 
\begin{equation}
|q_2| < \frac{ \ep(N+1-\ep)}{1+2\ep}.
\end{equation}
Provided that this condition holds, we can exchange the integral and the sum to obtain
\begin{equation}
\label{eq:lasty}
Y_{3+(2-N) n -b, b}^{\text{Higgs}} = q_1^n \oint_{C(-1)} \frac{\omega^b s(\omega)^n R^{-n} }{(1+2\omega)P(\omega)}.
\end{equation}
And now comes a pleasant surprise: the condition for convergence ensures that $\omega = \omega_+$ is 
enclosed by $C(-1)$, while $\omega = \omega_-$ remains outside of it, and so, since $R(\omega_\pm) =1$, 
\begin{equation}
\label{eq:lasty2}
Y_{3+(2-N) n -b, b}^{\text{Higgs}} = q_1^n \oint_{C(\omega_+)}  \frac{d\omega}{2\pi i}
                                      \frac{\omega^b s(\omega)^n }{(1+2\omega)P(\omega)},
\end{equation}
and for $n<0$ we precisely have the desired form for the correlators:
\begin{equation}
Y_{a,b}^{\text{Geometric}} = Y_{a,b}^{\text{Higgs}} + Y_{a,b}^{\text{Coulomb}}.  
\end{equation}

The contribution to the $n=0$ correlators is even more remarkable. The standard manipulation of the
instanton sum yields
\begin{equation}
\label{eq:lasty3}
Y_{3-b,b}^{\text{Higgs}} = Y_{3-b,b}^{0} + \sum_{m=0}^{\infty} \oint_{C(N)} \frac{d\omega}{2\pi i} 
                          \frac{\omega^b R^m}{(1+\omega)(-N+\omega)(-1-2\omega)},
\end{equation}
with $R$ as above, and $C(N)$ a small contour about $\omega = N$:  $\omega = N+ \ep e^{i\theta}$.
The constants $Y_{3-b,b}^0$ parametrize our ignorance of how to compute intersection
numbers on a non-compact variety.  Presumably, these are computed by an appropriate cohomolgy theory.
For uniform convergence we must have 
\begin{equation}
 |q_2| < \frac{ \ep (1-\ep)}{1+2N +\ep}.
\end{equation}
Carrying out the sum, we have
\begin{equation}
Y_{3-b,b}^{\text{Higgs}} = Y_{3-b,b}^{0} - \oint_{C(N)} \frac{d\omega}{2\pi i} 
                          \frac{\omega^b}{(1+2\omega)P(\omega)}.
\end{equation}
The condition for convergence ensures that, this time, $\omega_-$ is enclosed by $C(N)$,
while $\omega_+$ remains outside.  The crucial overall minus sign means that putting this together
with the Coulomb contribution, we find that the $Y_{3-b,b}$ are just constants, as predicted by
our discussion above. 

A similar analysis may be carried out in the other Mixed Phases.  In those phases all of the Coulomb 
vacua are reliable and contribute.  For $n<0$ there are no contributions from the gauge instantons of
the Higgs branch, while for $n=0$ the instanton sums cancel the Coulomb contribution up to the constants
$Y_{3-b,b}^0$.  It would be interesting to examine these constants in more detail, but whatever they are,
the resulting correlators are incompatible with quantum cohomology relations.

One final aspect of this example deserves mention---the disappearance of the semi-classical singularity
at $q_2 = 1/4.$  We know that semi-classical analysis of the Higgs branch in Mixed Phase B or C shows
this singularity.  We expect that analysis to be valid for small $|q_1|$. Of course, the discrete Coulomb
vacua also exist in this limit, and it is possible that the presence of these additional  Coulomb vacua 
washes out the singularity. It appears that this is so, at least in the topological theory.  

%%%%%%%%%%%%%%%%%%%%%%%%%%%%%%%%%%%%%%%%%%%%%%%%%%%%%%%%%%%%%%%
%%%%%%%%%%%%%%%%%%%%%%%%%%%%%%%%%%%%%%%%%%%%%%%%%%%%%%%%%%%%%%%%
\section{Discussion} \label{s:discuss}

We have found a simple algebraic formula for the Coulomb contribution to the $A$-model correlators in 
toric GLSMs.  We hope to have convinced the reader that this expression is conceptually
satisfying and computationally useful. We will now conclude with an outlook on some interesting
questions that remain.

\subsection{Some Observations on the Coulomb Vacua and the GLSM}

Our work is a simple application of the general principle of localization in TFTs. 
It has been known for a long time that in the Geometric Phases the path integral localizes 
onto the gauge instantons.  We have merely extended this result to the Mixed and Non-Geometric
Phases.  In models with a Non-Geometric Phase our result gives a surprisingly complete form for
the $A$-model correlators.  In models without such a phase we are restricted to genus
zero by our inability to compute the contribution from the Higgs vacua.

Perhaps the most surprising finding of our work is that for models without a Non-Geometric 
Phase the quantum cohomology relations may fail for a subset of the correlators which, in some
Mixed Phase, receive contributions from both Higgs and Coulomb vacua.  There are two simple ``proofs''
of quantum cohomology relations:  the first is a Geometric Phase analysis of the intersection numbers 
on the gauge instanton moduli spaces\cite{MP:summing,ME:glsmtach}, and the second is a Non-Geometric Phase analysis
given above.  The first argument is subtle when the phase corresponds to a non-compact
variety, and the second does not apply in the absence of a Non-Geometric Phase.  The example of
the last section illustrates that these problems are manifestations of the same failure of the quantum
cohomology relations in different phases.

We have not addressed computations of the correlators in phases where discrete Higgs-Coulomb vacua are 
present.  It would be interesting to understand the contributions from these vacua.  This exercise may 
well provide some new insights into the more mysterious aspects of GLSM physics, and it may provide us
with another set of phases where the computation of the correlators is made tractable.

Finally, we have derived our results in the context of traditional GLSMs.  It would be interesting
to extend our treatment to the recently much-discussed GLSMs corresponding to supermanifolds.

\subsection{A Pure Landau-Ginzburg Description?}
The contribution to the $A$-model correlators bears a striking resemblance to Vafa's result
on correlators in topological Landau-Ginzburg models \cite{VAFA:tlg}.  Vafa studied a topological
Landau-Ginzburg model with a superpotential $W(X)$, and he showed that the correlators are
given by
\begin{equation}
\label{eq:Vafa}
\la F(x) \ra = \Tr F(x) H^{g-1},
\end{equation}
where the trace is taken over the critical points of $W$, and each contribution is 
weighted by the Hessian of $W$, $H$, evaluated at that point.

While it is certainly not true that the GLSM model correlators are computed by a Landau-Ginzburg theory 
with $\Sigma_a$ as the fields and $\Wte$ as the superpotential, because of the remarkable similarity
between eqn.(\ref{eq:corresult}) and eqn.(\ref{eq:Vafa}), it is natural to wonder if {\em some} Landau-Ginzburg 
theory computes the same correlators.  It is easy to see that this is possible for at least some models. 
For example, a Landau-Ginzburg theory of a single field $\Sigma$ and superpotential 
$W = \frac{1}{6} \Sigma^6 - q\Sigma$ has the equation of motion $\sigma^5 = q$ and Hessian of $5\sigma^4$.  
Hence, according to eqn.(\ref{eq:Vafa}), this model will compute the same correlators as the GLSM with
target space $\P^4$.

It is clear that, in general, the requisite Landau-Ginzburg model will be much more complicated.  After all, 
the theories studied by Vafa can be constructed as relevant deformations of a free theory with an ultraviolet
$R$-symmetry with charges $Q_+(X) = Q_-(X) = 0$,  and all of the solutions to the {\em classical} equations of
motion $dW(x)=0$ are on the same footing.  It is tempting to suggest that the method of Hori and 
Vafa\cite{HV:mirsym}, which relies on dualizing the matter fields, yields this Landau-Ginzburg 
description.  It would be interesting to check whether this is the case.

\subsection{Coupling to Topological Gravity}
Another direction to pursue  is to couple the model to topological gravity in the spirit 
of \cite{W:tphase}.  The resulting theory would be an interesting topological string theory, 
where perhaps the $\lambda$-dependent poles we have found would find a natural interpretation.  
Furthermore, this model would, in principle, compute a much larger subset of non-trivial 
Gromov-Witten invariants.  Hopefully, the simplicity of our result for the correlators would 
persist to some extent in the string theory.

\section*{Acknowledgments}
It is a pleasure to thank C.~Haase , E.~Katz, G.~Moore, and S. Rinke for useful 
comments and conversations.  M.R.P. would like to thank the Perimeter 
Institute and the Theoretical High Energy Physics group at Rutgers University
for hospitality while some of this work was completed.  This article is 
based upon work supported in part by the National Science Foundation 
under Grants DMS-0074072 and DMS-0301476.  Any opinions, findings, and 
conclusions or recommendations expressed in this article are those of the 
authors and do not necessarily reflect the views of the National Science 
Foundation.

%\bibliographystyle{unsrt}
%\bibliography{big}

\end{document}